\documentclass{article}
\usepackage{ijcai26}

\usepackage{times}
\usepackage{soul}
\usepackage{url}
\usepackage[hidelinks]{hyperref}
\usepackage[utf8]{inputenc}
\usepackage[small]{caption}
\usepackage{graphicx}
\usepackage{amsmath}
\usepackage{amsthm}
\usepackage{booktabs}

\usepackage[switch]{lineno}

\usepackage{amssymb}
\usepackage{amsmath}
\usepackage{mathtools}
\usepackage{stmaryrd}
\usepackage{bm}

\usepackage{xcolor}      
\usepackage{graphicx}
\usepackage{booktabs}
\usepackage{tabularx}
\usepackage{makecell}
\usepackage{threeparttable}
\usepackage{multirow}
\usepackage{array}
\usepackage{colortbl}

\usepackage{subcaption}  

\usepackage[ruled,linesnumbered]{algorithm2e}

\usepackage{paralist}
\usepackage{enumitem}    
\usepackage{pifont}
\usepackage{microtype}
\usepackage{fontawesome} 
\usepackage{tcolorbox}
\usepackage{listings}

\usepackage{diagbox}
\usepackage{multicol}

\title{Improving Code Understanding in Large Language Models through Concept-Aware Consistency Learning}

\author{
    Xiaoning Ren$^1$ \and
    Qiang Hu$^2$ \and
    Wei Ma$^3$ \and
    Chongyang Liu$^1$ \and
    Yan Li$^1$ \and
    Yao Zhang$^2$ \and
    Lingxiao Jiang$^3$ \and
    Yongqiang Lyu$^{2,5}$ \and
    Yinxing Xue$^{4,*}$ 
    \affiliations
    $^1$University of Science and Technology of China
    \\
    $^2$Tianjin University
    \\
    $^3$Singapore Management University
    \\
    $^4$Institute of AI for Industries, Chinese Academy of Sciences  \thanks{Corresponding author.}
    \\
    $^5$ JCSS, Tsinghua University (INSC) - Science City (Guangzhou) Digital Technology Group Co., Ltd.
    \emails
    \{hnurxn, liyann\}@mail.ustc.edu.cn,
    \{qianghu, zzyy, lyuyq\}@tju.edu.cn,
    \{weima, lxjiang\}@smu.edu.sg,
    lcyyy9@gmail.com,
    yxxue@iaii.ac.cn
}

\begin{document}

\maketitle

\begin{abstract}
Although Large Language Models (LLMs) excel at code generation, recent research reveals that they exhibit an insufficient grasp of core programming concepts, such as data flow and control flow. This limitation undermines their robustness when encountering variations in these concepts in practice; however, effective solutions that explicitly target this gap remain limited.
To address this challenge, we propose \textsc{ProCURE}, a concept-aware consistency learning framework designed to enhance LLMs’ understanding of programming concepts. Specifically, \textsc{ProCURE} first performs automated concept-oriented code augmentation to construct a concept-aligned dataset covering  representative programming concepts. It then conducts concept-aware  fine-tuning, encouraging the model to capture fine-grained concept variations and learn appropriate generation behaviors under such variations via a novel concept-sensitive consistency loss.
To quantify programming concept understanding, we introduce the Concept Consistency Score (CCScore), defined as the proportion of correct generations preserved under concept variations.  A higher CCScore indicates a more profound understanding of programming concepts.
We evaluate \textsc{ProCURE} on four open-source LLMs across three widely used code generation benchmarks.  Experimental results show that \textsc{ProCURE} improves CCScore by an average of 17.9 points, demonstrating its effectiveness in addressing the programming concept understanding gap.
\end{abstract}

\section{Introduction}

Large language models (LLMs) have demonstrated promising performance in code generation tasks, marking a new paradigm for programming~\cite{zhang2023repocoder,schafer2023empirical,abdin2024phi}.
However, whether LLMs can truly understand code remains an open question. Such understanding is a prerequisite for users to reliably and confidently employ LLMs in code-related tasks, ensuring controllability and trustworthiness. Consequently, in-depth evaluation and enhancement of LLMs’ code understanding ability has emerged as a critical research direction for the community.

\begin{figure}[t]
	\centering
	\includegraphics[scale=0.44]{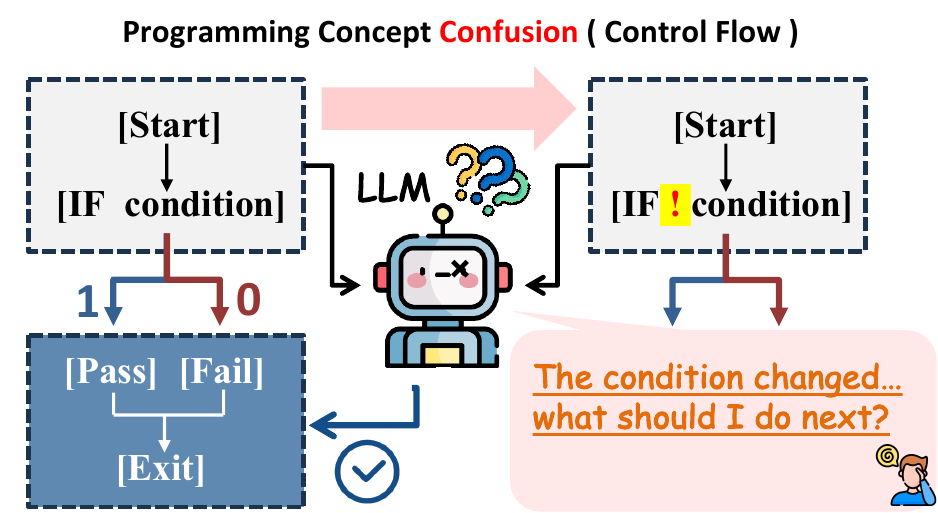}
	\caption{
		An illustrative example showing the limited understanding of LLMs to programming concept variations.
	}
	\label{fig:1}
    \vspace{-6pt}
\end{figure}

Existing works~\cite{ma2023lms,orvalho2025large} revealed that LLMs have the potential to grasp code syntax but do not understand the semantics  correctly. Even worse, recent research~\cite{hooda2024large} highlighted a fundamental limitation: LLMs lack sufficient understanding of \textit{Programming Concepts Predicates}\footnote{Throughout this paper, we use the term \textit{programming concepts} to denote programming concept predicates for brevity.} , including how variables are managed in memory, how execution flows across constructs, and how code segments combine to perform computations. 
As illustrated in Figure~\ref{fig:1}, the figure presents a representative failure case in which a model becomes confused when control-flow variations are introduced within the code prefix.

However, effective solutions that explicitly target programming concept understanding are urgently needed, yet largely absent in existing work.
Most prior approaches focus on improving robustness through surface-level perturbations and adversarial training~\cite{bielik2020adversarial,luo2025success}, or fine-tune models for other  objectives such as security~\cite{he2024instruction}, without explicitly modeling programming concepts or enforcing consistent behavior under concept-level variations.
To bridge this gap, we  propose a  novel framework \textbf{\textsc{ProCURE}}~( \underline{\textbf{Pro}}gramming \underline{\textbf{C}}oncept  \underline{\textbf{U}}nde\underline{\textbf{R}}standing \underline{\textbf{E}}nhancement),
which enhances programming concept understanding via concept-aware consistency learning.
\textsc{ProCURE} consists of two stages.
In the first stage, we construct a concept-aligned dataset through a fully automated, LLM-assisted pipeline, where concept-oriented perturbations are generated using enriched one-shot chain-of-thought prompts augmented with static code analysis signals.
In the second stage, we perform concept-aware fine-tuning with a concept-sensitive consistency loss that emphasizes concept-relevant tokens, encouraging the model to respond consistently to fine-grained programming concept variations while preserving functional correctness.

We conduct a comprehensive evaluation on open-source LLMs, including Llama3.1-8B, CodeLlama-13B, StarCoder-7B, and Mistral-7B, using three widely recognized benchmark datasets: HumanEval, MBPP, and CodeContests, which span a diverse range of programming challenges. 
Experimental results show that \textsc{ProCURE} improves CCScore by an average of 17.9 points, demonstrating its effectiveness in enhancing programming concept understanding and providing new insights for code augmentation and robust training in downstream code-related tasks. 
In summary, our contributions are outlined as follows:

\begin{compactitem}[$\bullet$]

	\item We propose \textsc{ProCURE}, a concept-aware consistency learning framework that enhances programming concept understanding in LLMs, which is, to the best of our knowledge, the first to explicitly target this problem.
	
	\item Extensive experiments show the effectiveness of \textsc{ProCURE}, improving CCScore by an average of 17.9 points.
	
	\item We  release \href{https://github.com/hnurxn/Counterfactual_Benchmark}{\textit{ConceptEval}}\footnote{https://github.com/hnurxn/Counterfactual\_Benchmark}, a new benchmark for evaluating programming concept understanding.

\end{compactitem}

\section{Preliminary}
\subsection{LLM-based Code Generation}

LLM-based code generation~\cite{jiang2024survey} formulates source code generation as an autoregressive process conditioned on natural language, source code, or both. 
Given an input token sequence \(T = [t_1, \dots, t_M]\), an LLM generates code by iteratively estimating the conditional probability of the next token:
\[
P(Y \mid T) = \prod_{i=1}^{N} P(y_i \mid T, y_1, \dots, y_{i-1}),
\]
where \(Y = [y_1, \dots, y_N]\) denotes the generated token sequence.
Among various code generation tasks, code completion is one of the most widely studied settings, where \(T\) consists of a functional instruction and a partial program.
Following ~\cite{hooda2024large}, we adopt code completion to evaluate LLMs’ understanding of programming concepts, as it directly tests whether models can appropriately adapt their generation to concept-level variations in the input.

\subsection{Programming Concepts}
\begin{figure*}[t]
	\centering
	\includegraphics[scale=0.8]{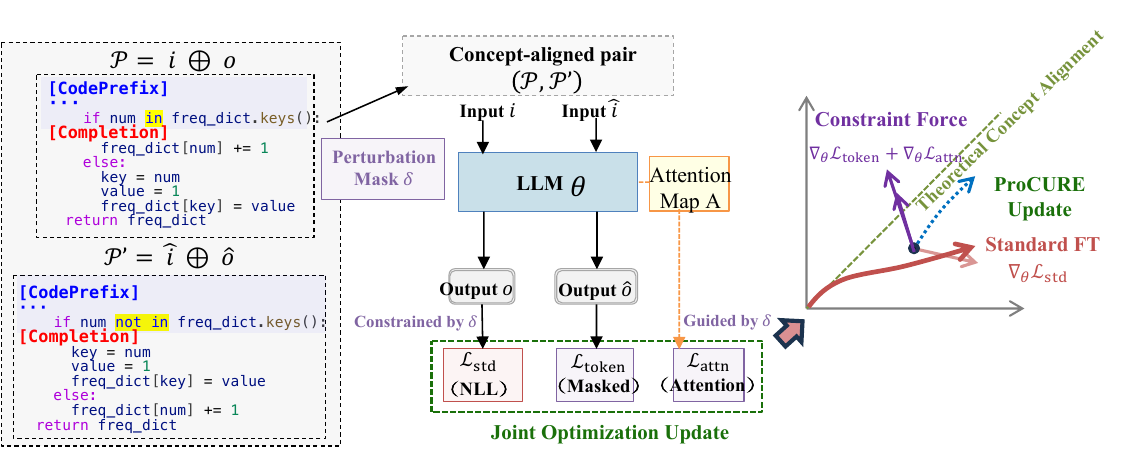}
	\caption{Overview of \textsc{ProCURE}. It first constructs a concept-aligned dataset through automated concept-oriented data augmentation, and then conducts joint optimization with a concept-sensitive consistency objective to enable concept-aware consistency learning.}
	\label{fig:2}
    \vspace{-6pt}
\end{figure*}

\noindent\textbf{Programming Concepts} refer to the properties of specific program elements such as variables, data values, data paths, and execution paths, considered from a holistic, program-wide perspective~\cite{hooda2024large,hoare1969axiomatic}. These concepts may describe, for example, the value range of a variable at a specific location, whether control flow can reach one function location from another, or whether a value assigned to a variable may later be used elsewhere (i.e., data-flow or control-flow).
Prior work identifies four major themes —control flow~\cite{allen1970control,yang2015static}, data flow~\cite{nilsson2009declarative,fosdick1976data}, identifier names~\cite{lin2008evaluation}, and data types~\cite{allamanis2020typilus,dart1992efficient}—and proposes a representative set of programming concepts that cover these themes:

\begin{compactitem}[$\bullet$]
\item \textbf{Control Flow} describes the execution order of statements governed by control structures. Perturbing a control condition (e.g., in an if statement) requires adjusting the corresponding execution path.
\item \textbf{Data Flow} ensures  variable references are in scope and live at their use sites, preserving correct data dependencies and value propagation.
\item \textbf{Data Types} constrain the set of valid values and operations for variables; type correctness must be maintained to ensure semantic soundness.
\item \textbf{Identifier Naming} refers to the symbolic names of variables. While not affecting semantics, consistent renaming is necessary to preserve referential correctness.
\item \textbf{Statement Order} captures cases where independent statements without mutual data or control dependencies can be reordered without affecting program behavior.

\end{compactitem}
In this work, we follow the problem formulation of prior study~\cite{hooda2024large} that identifies the programming concept understanding gap in LLMs. Specifically, we focus on weakly typed programming languages (e.g., Python), where type information is either implicit or dynamically resolved at runtime. Consequently, our analysis centers on the four programming concepts introduced above, excluding data types.

\noindent\textbf{What It Means to Understand Programming Concepts?} 
Understanding a programming concept  means that when a concept-oriented change occurs at a specific program location, an appropriate response can be made based on the underlying programming logic to preserve the program’s intended functionality.
Let \(P = \langle l_{1}, l_{2}, \dots, l_{n} \rangle\) be a program of \(n\) lines, with its overall functionality denoted by \(\llbracket P \rrbracket = F\).  
Let \(\mathcal{E} = \{e_{1}, \dots, e_{m}\}\) be the set of $m$ program elements.  
Suppose a program element \(e_j \in \mathcal{E}\) is modified at line \(k\) (\(1 \leq k \leq n\)), resulting in an intermediate program \(P \oplus \Delta(e_j)\).
Understanding a programming concept implies that an appropriate transformation \(T_{\mathcal{C}(e_j)}\) can be applied, which rewrites only the lines within the impact region \(\mathcal{C}(e_j) \subseteq \{l_{k+1}, \dots, l_n\}\), such that:
\begin{equation}
\label{eq:ple-preserve}
\llbracket
T_{\mathcal{C}(e_{j})}
\bigl(P \oplus \Delta(e_{j})\bigr)
\rrbracket = 
\llbracket
P
\rrbracket
\end{equation}
In Equation~\eqref{eq:ple-preserve}, the transformation \(T_{\mathcal{C}(e_j)}\), guided by the semantics of the corresponding programming concept, compensates for the effect of \(\Delta(e_j)\) by refactoring only the affected lines, thereby preserving the original functionality \(F\).

When applied to LLMs, this definition implies that given a functional instruction and a code prefix, the model should adapt its completion to concept-related changes in the prefix while preserving the intended functionality.
However, existing LLMs exhibit limited understanding of such programming concepts, and effective solutions remain lacking.
Addressing this gap constitutes the primary motivation of our work.

\section{Framework: \textsc{ProCURE} }

\subsection{Overview  }

As illustrated in Figure~\ref{fig:2}, we first develop an automated concept-oriented perturbation engine that systematically applies transformations to input programs with respect to specific programming concepts.
Each transformation generates a perturbed variant paired with the original program, and all such pairs collectively form a concept-aligned dataset.
Built upon this dataset, we perform joint optimization during fine-tuning, where the standard training objective is augmented with a concept-sensitive consistency objective that emphasizes concept-relevant regions.
This optimization encourages the model to preserve functional correctness while appropriately responding to concept-level variations, thereby enabling LLMs to internalize programming concepts and exhibit consistent behavior under concept-oriented changes.

\subsection{Concept-oriented Perturbation Engine}

\begin{figure}[t]
	\centering
	\includegraphics[scale=0.23]{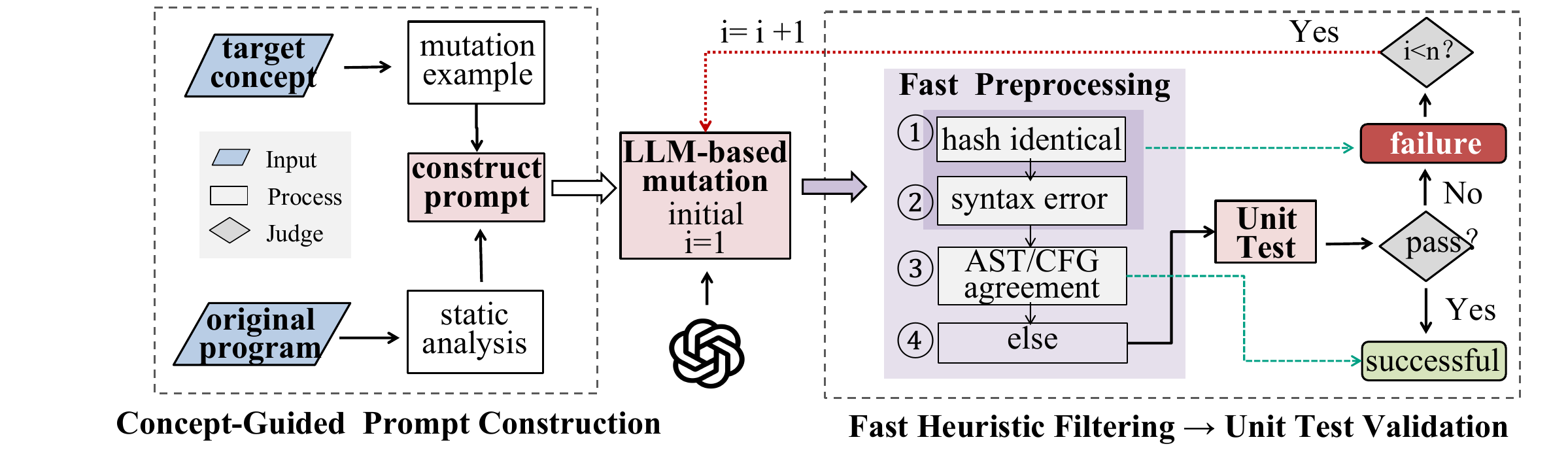}
	\caption{The overall pipeline for concept-oriented perturbation engine, consisting of prompt construction  and result validation.
	}
	\label{fig:3}
	\vspace{-4mm}
\end{figure}

As illustrated in Figure~\ref{fig:3}, we present the pipeline of LLM-assisted Concept-Oriented Perturbation Engine.
It consists of two core components: (1) efficient and effective prompt construction for concept-oriented perturbation generation, and (2) automatic validation of the generated perturbations to ensure semantic correctness.
By repeatedly applying this pipeline to the original programs, we construct a high-quality concept-aligned dataset for subsequent concept-aware fine-tuning.

\noindent \textbf{Concept-Guided  Prompt Construction.}
The perturbation engine is driven by a set of concept-specific perturbation strategies that modify targeted structural aspects of a program while preserving function equivalence.
Following prior work~\cite{hooda2024large,ma2024speceval}, we consider five representative strategies: If-Else Flip, Def-Use Break, Independent Swap, Name Random, and Name Shuffle.
To guide the LLM toward generating valid and concept-aligned perturbations, we perform concept-specific static analysis on the input program and incorporate the extracted signals into the prompt.
We further augment the prompt with chain-of-thought reasoning and a one-shot example to improve generation reliability.
Details of the perturbation rules, illustrative examples, and the complete prompt template are provided in the Supplementary Material.

\noindent \textbf{Results Validation.}
After the LLM generates a perturbed candidate, we apply an automated validation procedure to determine whether it constitutes a valid function-preserving transformation.
The procedure follows a two-stage design, consisting of a fast, execution-free filtering stage followed by unit test validation.

\noindent\textit{Fast Heuristic Filtering.}
We first apply lightweight checks to quickly eliminate invalid or redundant candidates without executing the program.
Specifically, we discard candidates that are identical to the original program based on hash comparison, or that contain syntax errors detected via AST parsing.
For the remaining candidates, we examine structural differences: if the abstract syntax trees are identical, or if the control flow graphs remain unchanged despite syntactic variations, the candidate is accepted as a valid perturbation at this stage.
\noindent\textit{Unit Test Validation.}
Candidates that pass heuristic filtering but whose semantic equivalence remains uncertain are further validated through unit test execution.
Each original program is associated with a set of functional test cases, and a perturbed variant is accepted only if it passes all tests.
Candidates failing any test are rejected.

\noindent \textbf{Concept-Aligned Dataset Construction.}
All validated perturbations are paired with their corresponding original programs to form concept-aligned program pairs.
For each pair, we annotate the transformation locations at the token level on the perturbed variant using \textit{difflib}\footnote{\url{https://docs.python.org/3/library/difflib.html}}, explicitly identifying tokens modified by the concept-oriented perturbation.
The resulting collection of annotated pairs constitutes our concept-aligned dataset, which serves as training data for concept-aware fine-tuning and as the basis for constructing an evaluation benchmark.

\subsection{Concept-Aware  Fine-Tuning}
\label{finetuning}
\begin{algorithm}[t]
\caption{Concept-Aware Fine-Tuning}
\label{alg:concept_aware_ft}
\KwIn{Dataset \(\mathcal{D}_{\text{combine}}\), learning rate \(\eta\), number of epochs \(E\)}
\KwOut{Fine-tuned model parameters \(\theta_{\text{FT}}\)}

Initialize model parameters \(\theta\);

\For{$e = 1$ \KwTo $E$}{
	Shuffle \(\mathcal{D}_{\text{combine}}\);
	
	\ForEach{mini-batch \(\mathcal{B}\) sampled from \(\mathcal{D}_{\text{combine}}\)}{
		Partition \(\mathcal{B}\) into original samples \(\mathcal{B}_{\text{ori}}\) and concept-aligned samples \(\mathcal{B}_{\text{con}}\);
		
		Compute the standard NLL loss
		\(\mathcal{L}_{\text{std}}\) on \(\mathcal{B}_{\text{ori}}\);

		\ForEach{concept-aligned sample \((\mathbf{i}^{c}, \mathbf{o}^{c}) \in \mathcal{B}_{\text{con}}$}{
			Compute masked token loss \(\mathcal{L}_{\text{token}}\);
			
			Compute attention-guided loss \(\mathcal{L}_{\text{attn}}\);
		}
		
		Compute the overall batch loss:
				\(
		\mathcal{L}_{\text{batch}} =
		\mathcal{L}_{\text{std}} +
		\mathcal{L}_{\text{token}} +
		\mathcal{L}_{\text{attn}} ;
		\)
		
		Update model parameters:
		\(\theta \leftarrow \theta - \eta \cdot \nabla_{\theta} \mathcal{L}_{\text{batch}}\);
	}
}
\Return{\(\theta_{\text{FT}}\)}
\end{algorithm}

We propose a concept-aware  fine-tuning strategy that jointly optimizes standard task learning and programming concept understanding.
The core idea is to integrate concept-aware data sampling with a concept-sensitive loss, enabling the model to learn task-specific generation behavior while consistently responding to programming concept variations.

\noindent\textbf{Concept-Aware Data Sampling.}
Given an original dataset
\(\mathcal{D}^{\text{ori}} = \{(\mathbf{i}_j, \mathbf{o}_j)\}\),
where each input \(\mathbf{i}_j = (h, x)\) consists of a functional instruction \(h\) and a code prefix \(x\), and \(\mathbf{o}_j = y\) denotes the corresponding target completion,
and a set of programming concepts \(\mathbb{C} = \{c_1, c_2, \dots, c_n\}\),
we construct a concept-aligned dataset
\(\mathcal{D}^{\text{con}} = \{(\mathbf{i}_j^{c_k}, \mathbf{o}_j^{c_k})\}\)
by applying concept-oriented perturbations with respect to each \(c_k \in \mathbb{C}\).
Both the input and output are transformed in a concept-consistent manner to preserve the  program functionality.

We define the combined training dataset as:
\begin{align*}
	\resizebox{1\linewidth}{!}{$
		\mathcal{D}_{\text{combine}} =
		\Big\{
		\big(
		(\mathbf{i}_j, \mathbf{o}_j),
		(\mathbf{i}_j^{c_1}, \mathbf{o}_j^{c_1}),
		\dots,
		(\mathbf{i}_j^{c_n}, \mathbf{o}_j^{c_n})
		\big)
		\;\Big|\;
		(\mathbf{i}_j, \mathbf{o}_j) \in \mathcal{D}^{\text{ori}}
		\Big\}
		$}.
\end{align*}

Each entry in \(\mathcal{D}_{\text{combine}}\) consists of one original example and its associated concept-aligned variants, all sharing the same underlying functionality.
To ensure joint exposure to both sample types during training, we adopt a structured batch construction strategy.
Specifically, each mini-batch \(\mathcal{B}\) is formed as:
\[
\mathcal{B} =
\bigcup_{j=1}^{|B| / (n+1)}
\left\{
(\mathbf{i}_j, \mathbf{o}_j),
(\mathbf{i}_j^{c_1}, \mathbf{o}_j^{c_1}),
\dots,
(\mathbf{i}_j^{c_n}, \mathbf{o}_j^{c_n})
\right\},
\]
where \(|B|\) denotes the total batch size.
This design ensures that original and concept-aligned samples are jointly optimized within each parameter update.

\noindent \textbf{Concept-Sensitive Loss Function.}
Given the mixed nature of the training data, we employ a joint optimization objective composed of two complementary loss components.
Original samples focus on learning correct task behavior, while concept-aligned samples explicitly encourage sensitivity to programming concept variations.

\noindent \textit{Standard Loss.}
For each original sample \((\mathbf{i}, \mathbf{o}) \in \mathcal{D}^{\text{ori}}\), we apply the standard negative log-likelihood (NLL) loss:
\[
\mathcal{L}_{\text{std}}(\mathbf{i}, \mathbf{o}) =
-\sum_{t=1}^{|\mathbf{o}|}
\log P(\mathbf{o}_t \mid \mathbf{o}_{<t}, \mathbf{i}).
\]

\noindent \textit{Concept-Sensitive Loss.}
Let \(\mathbf{o}\) and \(\mathbf{o}^{c}\) denote the original and concept-aligned outputs, respectively.
To localize the effect of concept-oriented perturbations, we construct a binary mask
\(\boldsymbol{\delta} \in \{0,1\}^{|\mathbf{o}^{c}|}\),
where \(\delta_j = 1\) indicates that token \(j\) is affected by the perturbation.
We design two complementary loss terms to guide concept-aware learning:

\begin{compactitem}[$\bullet$]
	\item  Masked Token Loss.
	This term emphasizes prediction accuracy on concept-altered regions:
	\[
	\mathcal{L}_{\text{token}}(\mathbf{i}^{c}, \mathbf{o}^{c}) =
	-\sum_{j=1}^{|\mathbf{o}^{c}|}
	\delta_j
	\log P(\mathbf{o}_j^{c} \mid \mathbf{o}^{c}_{<j}, \mathbf{i}^{c}).
	\]
	
	\item  Attention-Guided Loss.
	To further encourage structural awareness, we regularize the decoder attention so that tokens generated after a perturbation attend more strongly to the altered regions.
    Given the decoder attention matrix
    \(\mathbf{M}^{\text{attn}} \in \mathbb{R}^{|\mathbf{o}^{c}| \times |\mathbf{o}^{c}|}\),
    we define:
    \[
    \mathcal{L}_{\text{attn}}(\mathbf{i}^{c}, \mathbf{o}^{c}) =
    -\lambda
    \sum_{j=1}^{|\mathbf{o}^{c}|}
    \sum_{k<j}
    \delta_k \cdot \left|M^{\text{attn}}_{jk}\right|,
    \]
    where \(\lambda > 0\) controls the strength of the attention regularization.
\end{compactitem}

\noindent \textbf{Training Summary.}
During training, each mini-batch contains both original and concept-aligned samples to enable joint optimization.
Specifically, original samples are optimized using the standard negative log-likelihood loss, while concept-aligned samples are trained with a concept-sensitive objective composed of the masked token loss and the attention-guided consistency loss.
The overall batch loss is computed as:
\[
\mathcal{L}_{\text{batch}} =
\sum_{(\mathbf{i}, \mathbf{o}) \in \mathcal{B}_{\text{ori}}}
\mathcal{L}_{\text{std}}
\;+\;
\sum_{(\mathbf{i}^{c}, \mathbf{o}^{c}) \in \mathcal{B}_{\text{con}}}
\left(
\mathcal{L}_{\text{token}} + \mathcal{L}_{\text{attn}}
\right),
\]
and model parameters are updated via gradient descent.
The complete training procedure is summarized in Algorithm~\ref{alg:concept_aware_ft}.
At each epoch, the combined dataset is shuffled and processed in mixed mini-batches containing both original and concept-aligned samples. Original samples are optimized using the standard negative log-likelihood loss to preserve functional correctness, while concept-aligned samples are trained with a concept-sensitive objective composed of the masked token loss and the attention-guided  loss.
These loss components are jointly optimized within each parameter update via gradient descent, enabling the model to balance task-level performance and programming concept sensitivity.
After all training epochs, the fine-tuned model parameters \(\theta_{\text{FT}}\) are obtained.

\section{Experiment Evaluation}

\subsection{Experimental Setup}
\label{setup}
\noindent\textbf{Models.}
We evaluate four state-of-the-art open-source LLMs, representing both general-purpose and code-specialized paradigms. For general-purpose LMs, we select LLaMA3.1-8B~\cite{touvron2024llama3} and Mistral-7B~\cite{jiang2023mistral}, both trained on large-scale web and code corpora. For coding LMs, we adopt StarCoder-7B~\cite{li2023starcoder} and CodeLLaMA-13B~\cite{roziere2023code}, which are pretrained on permissively licensed GitHub code and optimized for program synthesis tasks. 

\noindent\textbf{Benchmarks.}
We evaluate models on three widely-used code generation benchmarks: HumanEval~\cite{chen2021evaluating}, MBPP~\cite{austin2021mbpp}, and CodeContests~\cite{li2022competition}. HumanEval consists of 164 Python problems with unit tests for functional correctness. MBPP includes 1,000 beginner-level programming tasks, and CodeContests provides a challenging set of real-world algorithmic problems from competitive programming platforms. Together, these benchmarks span a broad range of task difficulties, algorithmic patterns, and prompt formulations. 
Building upon these benchmarks, we further construct a concept-variation benchmark for evaluating programming concept understanding under controlled concept-level perturbations. 

\noindent\textbf{Evaluation Metrics.}
We define a binary attribution function \( A: \mathcal{H} \times \mathcal{X} \times \mathcal{Y} \rightarrow \{0, 1\} \), where \(\mathcal{H}\) denotes the set of natural language instructions, \(\mathcal{X}\) the input space, and \(\mathcal{Y}\) the output space. \(A(h, x, y)\) is 1 if the model's output satisfies the instruction, and 0 otherwise. This attribution function is used to compute the following metrics:

\begin{compactitem}[$\bullet$]
\item \textbf{Pass@k.}  
Pass@k~\cite{abdin2024phi,luo2023wizardcoder,wei2023magicoder} measures the fraction of problems where at least one of \(k\) generated outputs satisfies the instruction:
\[
\text{Pass@}k = \frac{1}{n} \sum_{i=1}^{n} \left( 1 - \prod_{j=1}^{k} (1 - A(h_i, x_i, y_i^j)) \right)
\]
where \(n\) is the number of problems, and \(y_i^j\) is the \(j\)-th output for problem \(i\).

\item \textbf{Concept Consistency Score (CCScore).}  
Inspired by prior work~\cite{hooda2024large}, we quantify programming concept understanding using the Concept Consistency Score (CCScore), which measures the proportion of correct generations preserved under concept-level variations.
Given an input $x$ and its set of concept-perturbed variants $\{x^{c_k}\}_{k=1}^{K}$ ($K \leq 5$), the model is considered consistent on $x$ if it produces identical attribution outcomes across the original input and all its variants, i.e., $A(h, x, M(h, x)) = A(h, x^{c_k}, M(h, x^{c_k}))$ for all $k$.

The CCScore is computed as:
\begin{equation*}
	\resizebox{.99\linewidth}{!}{
	$\frac{\sum\limits_{(h, x)}
		\mathbb{I}\!\left[
		\forall k,\;
		A(h, x, M(h, x)) = A(h, x^{c_k}, M(h, x^{c_k}))
		\right]}
	{
		\sum\limits_{(h, x)}
		\mathbb{I}\!\left[
		(A(h, x, M(h, x)) + \sum_{k=1}^{K} A(h, x^{c_k}, M(h, x^{c_k})) > 0
		\right].
	}$}
\end{equation*}

Inputs are excluded only when the original output and all corresponding variant outputs are invalid, since such failures primarily indicate general model capacity  rather than concept understanding.
A higher CCScore indicates stronger robustness to concept-level variations.
\end{compactitem}

\noindent\textbf{Environment.}
All experiments are conducted on a single NVIDIA L40 GPU with 46GB of memory using PyTorch and the HuggingFace Transformers library.

\noindent\textbf{Hyperparameters.}
For automated concept-oriented perturbation, we use GPT-5 with temperature 1, generating up to 5 attempts per sample for valid outputs. For concept-aware fine-tuning, we optimize models using AdamW with a learning rate of  $5 \times 10^{-6}$, sequence length of 2048, batch size of 16, and a concept-sensitive loss with \(\lambda = 0.1\) over 5 epochs. During evaluation, we use top-p = 0.95 and max\_tokens = 512, with temperature set to 0.2 for Pass@1 and CCScore (averaged over 10 runs) and 0.6 for Pass@5, following prior work~\cite{chen2021evaluating,nijkamp2022conversational}.

\noindent \textbf{Baselines.}
Since there is currently no fine-tuning method explicitly designed to enhance \emph{programming concept understanding}, we adopt standard supervised fine-tuning (\textit{std-SFT}) and several representative code augmentation–based fine-tuning approaches as baselines. Specifically, we include: 
(1) a diversity-oriented code transformation baseline \textit{SA-AFT}~\cite{chen2023evaluating} that applies syntax- and structure-level transformations to increase data diversity; 
(2) a counterfactual instruction-tuning baseline  \textit{CTF-Code}~\cite{luo2025success}, which augments instructions via fine-grained counterfactual variations in problem descriptions; and 
(3) a robustness-oriented adversarial fine-tuning baseline \textit{CodeFort}~\cite{zhang2024codefort}, which incorporates semantic-preserving adversarial perturbations with robust training objectives.

\subsection{Dataset Experimental Results}

\begin{table*}[t]
\centering
\resizebox{0.8\linewidth}{!}{
\begin{tabular}{lcccccc}
\toprule
\textbf{Dataset} & \textbf{If-else Flip} & \textbf{Def-use Break} & \textbf{Independent Swap} & \textbf{Name Random} & \textbf{Name Shuffle} & \textbf{Avg} \\
\midrule
HumanEval       & 24 / 24 & 37 / 37 & 144 / 145 & 141 / 145 & 145 / 145 & 98.99\% \\
MBPP            & 198 / 200 & 179 / 182 & 957 / 972 & 924 / 972 & 946 / 972 & 97.15\% \\
CodeContests    & 3796 / 3821 & 4895 / 5564 & 6980 / 7004 & 7183 / 7221 & 6864 / 7221 & 96.39\% \\
\bottomrule
\end{tabular}
}
\caption{Perturbation success rates across datasets and perturbation types.}
\label{tab:cf_success}
\end{table*}
\begin{table*}[t]
\centering
\resizebox{0.8\linewidth}{!}{
\begin{tabular}{lcccccc}
\toprule
\textbf{Dataset} & \textbf{If-else Flip} & \textbf{Def-use Break} & \textbf{Independent Swap} & \textbf{Name Random} & \textbf{Name Shuffle} & \textbf{Avg (All)} \\
\midrule
HumanEval       
& 1.10~\textbar{}~\phantom{1}673 
& 1.22~\textbar{}~\phantom{1}803 
& 1.11~\textbar{}~\phantom{1}625 
& 1.44~\textbar{}~\phantom{1}841 
& 1.35~\textbar{}~\phantom{1}704 
& 1.24~\textbar{}~\phantom{1}729 \\
MBPP            
& 1.13~\textbar{}~\phantom{1}833 
& 1.30~\textbar{}~1001 
& 1.05~\textbar{}~\phantom{1}657 
& 1.14~\textbar{}~\phantom{1}718 
& 1.09~\textbar{}~\phantom{1}637 
& 1.14~\textbar{}~\phantom{1}769 \\
CodeContests    
& 1.15~\textbar{}~1347 
& 1.98~\textbar{}~2384 
& 1.06~\textbar{}~1177 
& 1.08~\textbar{}~1250 
& 1.59~\textbar{}~1969 
& 1.37~\textbar{}~1625 \\
\bottomrule
\end{tabular}
}
\caption{Average attempts and token costs across datasets and perturbation types.}
\label{tab:2}
\end{table*}

\begin{table}[t]
	\centering
	\resizebox{0.9\linewidth}{!}{
		\begin{tabular}{lccccc}
			\toprule
			& \multicolumn{2}{c}{\textbf{Perplexity (PPL)}} 
			& \multicolumn{3}{c}{\textbf{Human Annotation}} \\
			\cmidrule(lr){2-3} \cmidrule(lr){4-6}
			\textbf{Dataset}
			& \textbf{Original} & \textbf{Perturbed}
			& \textbf{Succ.} & \textbf{Fail.} & \textbf{Cohen’s $\kappa$} \\
			\midrule
			HumanEval    
			& 2457.73 & 1410.81
			& 99.5\% & 100\% & 0.995 \\
			MBPP         
			& 3117.99 & 1071.92
			& 99.0\% & 100\% & 0.990 \\
			CodeContests 
			& 1170.14 & 784.87
			& 98.5\% & 99.0\% & 0.975 \\
			\bottomrule
		\end{tabular}
	}
    	\caption{Quality Analysis of the Constructed Dataset.}
	\label{tab:dataset_quality}
\end{table}

\begin{table}[t]
	\centering
	\resizebox{0.9\linewidth}{!}{
		\begin{tabular}{lcccccc}
			\toprule
			& \multicolumn{3}{c}{\textbf{Exact Match}} 
			& \multicolumn{3}{c}{\textbf{Near Duplicate}} \\
			\cmidrule(lr){2-4} \cmidrule(lr){5-7}
			\textbf{Source} 
			& \textbf{HumanEval} & \textbf{MBPP} & \textbf{CodeContests}
			& \textbf{HumanEval} & \textbf{MBPP} & \textbf{CodeContests} \\
			\midrule
			\textbf{HumanEval}    
			& 0 & 0 & 0
			& 2 & 0 & 0 \\
			\textbf{MBPP}         
			& 0 & 7 & 0
			& 0 & 45 & 0 \\
			\textbf{CodeContests} 
			& 0 & 0 & 64
			& 0 & 0 & 101 \\
			\bottomrule
		\end{tabular}
	}
	\caption{Exact and Near-Duplicate Detection Results (pair counts).}
	\label{tab:dup}
\end{table}

\noindent \textbf{Effectiveness and Efficiency of Dataset Construction.}
Table~\ref{tab:cf_success} summarizes the success rates of concept-oriented perturbation generation across datasets and perturbation types.
Overall, the proposed pipeline achieves a high success rate of 97.51\%, with dataset-level averages of 98.99\%, 97.15\%, and 96.39\% on HumanEval, MBPP, and CodeContests, respectively.
Among perturbation strategies, \textit{If-Else Flip} yields the highest success rate (99.33\%), while \textit{Name Shuffle} performs worst, particularly on CodeContests (95.06\%).
Table~\ref{tab:2} reports the efficiency of the perturbation process.
Across all settings, valid perturbations are obtained with fewer than two attempts on average (1.24 overall), and the average token cost is 1,041 tokens per sample.
CodeContests incurs higher costs due to its greater program complexity.

\noindent \textbf{Dataset Quality.}
Before fine-tuning, we systematically assess the quality of the constructed concept-aligned dataset \(\mathcal{D}^{\text{con}}\) from both automatic statistics and human evaluation, ensuring that the generated perturbations are natural and semantically reliable.
Code naturalness reflects the readability and predictability of generated programs.
We measure naturalness using perplexity (PPL), computed by a pre-trained CodeGPT model (\textit{microsoft/CodeGPT-small-py}), where lower values indicate more natural code.
As shown in Table~\ref{tab:dataset_quality}, the perturbed programs consistently achieve lower PPL than the original programs across all benchmarks (e.g., decreasing from 2457.73 to 1410.81 on HumanEval).
This trend suggests that concept-oriented perturbations preserve, and in some cases improve, the linguistic regularity of code while introducing structural variations.
To further verify semantic validity, we conduct a human evaluation on 400 randomly sampled program pairs, evenly split between valid and invalid perturbations.
Two PhD-level annotators independently assess whether each perturbed program correctly reflects the intended programming concept while preserving functionality.
Table~\ref{tab:dataset_quality} reports high inter-annotator agreement across all datasets, with Cohen’s $\kappa$ exceeding 0.975, indicating near-perfect consistency.
These results confirm that our automated perturbation pipeline produces high-quality, semantically valid samples suitable for concept-aware fine-tuning.

\noindent \textbf{Data Decontamination.}
Data leakage poses a significant threat to the validity of LLM evaluation.
To mitigate this risk, we conduct a comprehensive decontamination analysis at two granularities:
(i) \textit{Exact Match} detection via hash comparison, and
(ii) \textit{Near-Duplicate Detection} using MinHash with locality-sensitive hashing (LSH) over 5-gram token shingles, with a similarity threshold of $\tau=0.8$~\cite{kocetkov2022stack}.
The detection results are summarized in Table~\ref{tab:dup}.
No cross-dataset leakage is observed among HumanEval, MBPP, and CodeContests.
However, we identify limited internal redundancy within individual datasets.
Specifically, CodeContests contains 64 exact duplicates and 101 near-duplicate pairs, while MBPP exhibits minor redundancy (7 exact and 45 near duplicates), and HumanEval shows only 2 near-duplicate pairs.
To ensure a fair and uncontaminated evaluation, we apply a strict sanitization procedure, retaining only a single representative from each detected duplicate cluster and removing all remaining instances.

\noindent \textbf{Dataset Summary.} Through the above process, we construct a concept-aligned dataset comprising 33,203 perturbed programs of varying complexity, derived from 8,285 original samples, with each sample associated with 1–5 concept-oriented variants.
This dataset is directly applicable to concept-aware fine-tuning.
Moreover, by integrating the CCScore evaluation protocol into the data construction pipeline, we further build and release the first open benchmark, \href{https://sites.google.com/view/procurecode/benchmark}{\textit{ConceptEval}}, for evaluating programming concept understanding in LLMs.

\subsection{Fine-tuning Experimental Results}

\begin{table*}[t]
	\centering
	\small
	\setlength{\tabcolsep}{3.5pt}
	\begin{tabular}{l ccc ccc ccc ccc}
		\toprule
		\multirow{2}{*}{\textbf{Method}} & 
		\multicolumn{3}{c}{\textbf{Mistral-7B}} & 
		\multicolumn{3}{c}{\textbf{Llama3.1-8B}} & 
		\multicolumn{3}{c}{\textbf{StarCoder-7B}} & 
		\multicolumn{3}{c}{\textbf{CodeLlama-13B}} \\
		\cmidrule(lr){2-4} \cmidrule(lr){5-7} \cmidrule(lr){8-10} \cmidrule(lr){11-13}
		& pass@1 & pass@5 & CCScore 
		& pass@1 & pass@5 & CCScore 
		& pass@1 & pass@5 & CCScore 
		& pass@1 & pass@5 & CCScore \\
		\midrule
		
		Base (w/o FT)
		& 43.1 & 49.8 & 31.1
		& 51.2 & 58.5 & 32.4
		& 44.6 & 48.2 & 38.1
		& 41.5 & 45.3 & 24.8 \\
		
		std-SFT
		& 49.2 & 56.1 & 33.2
		& 55.3 & 62.4 & 34.7
		& 48.4 & 53.9 & 38.9
		& 44.8 & 50.9 & 26.6 \\
		
		SA-AFT
		& \underline{61.2} & \textbf{74.6} & 39.6
		& \underline{64.3} & \underline{74.8} & 44.1
		& 62.1 & \textbf{72.9} & 43.1
		& 52.9 & 69.1 & 34.2 \\
		\addlinespace[0.3ex]
		
		CTF-Code
		& 56.3 & 66.9 & 36.4
		& 59.2 & 69.5 & 40.3
		& 55.1 & 65.8 & 39.1
		& \underline{53.6} & \underline{68.2} & 32.4 \\
		
		CodeFort
		& 59.8 & 70.4 & \underline{42.0}
		& 61.6 & 72.6 & \underline{46.2}
		& 58.9 & 68.7 & \underline{47.3}
		& 51.7 & 64.9 & \underline{36.1} \\
		
		\textbf{Ours}
		& \textbf{65.4} & \underline{73.7} & \textbf{46.4}
		& \textbf{67.8} & \textbf{76.5} & \textbf{50.9}
		& \textbf{65.0} & \underline{72.5} & \textbf{59.2}
		& \textbf{55.4} & \textbf{73.0} & \textbf{41.5} \\
		
		\bottomrule
	\end{tabular}
	\caption{Performance comparison of various fine-tuning techniques across four models, where Ours refers to ProCURE. The best results are bolded, and second-best are underlined.}
	\label{tab:main-results}
\end{table*}

\begin{table}[t]
	\centering
	\small
	\setlength{\tabcolsep}{6pt}
	\begin{tabular}{lccc}
		\toprule
		\textbf{Method} & \textbf{pass@1} & \textbf{pass@5} & \textbf{CCScore} \\
		\midrule
		w/o $\mathcal{L}_{\text{token}}$      
		& 57.4 & 65.9 & 39.9 \\
		w/o $\mathcal{L}_{\text{attn}}$       
		& \underline{61.5} & \underline{71.1} & \underline{47.6} \\
		w/o $\mathcal{L}_{\text{token}}+\mathcal{L}_{\text{attn}}$ 
		& 57.2& 65.3 & 38.3 \\
		\textbf{Ours}                         
		& \textbf{63.4} & \textbf{73.9} & \textbf{49.6} \\
		\bottomrule
	\end{tabular}
	\caption{Ablation results averaged over four models.}
	\label{tab:ablation_avg}
\end{table}


Based on the validated concept-aligned dataset, we evaluate the effectiveness of different fine-tuning strategies in improving both task performance and programming concept understanding. We randomly split the dataset into two equal parts, using 50\% for fine-tuning and 50\% for testing. For all baseline methods, data augmentation is applied only to the original training samples, and the number of augmented instances is strictly controlled to ensure that all methods are trained with the same total number of training samples, resulting in comparable training cost.
During evaluation, Pass@1 and Pass@5 are computed on the original test samples to measure task-solving capability, while the CCScore is computed on paired original and concept-aligned test samples to assess robustness under programming concept variations.

\noindent\textbf{Overall Results.}
Table~\ref{tab:main-results} summarizes the performance of all methods across four code LLMs. Overall, ProCURE consistently achieves the highest CCScore across all evaluated models, with an average improvement of 17.9 percentage points, demonstrating its effectiveness in strengthening programming concept understanding. In terms of task performance, ProCURE achieves consistently strong Pass@1 and Pass@5 results, ranking as the best or second-best across all settings. In contrast, standard supervised fine-tuning (std-SFT) yields noticeable gains in Pass@k but only marginal improvements in CCScore, indicating that while conventional fine-tuning improves task accuracy, it provides limited supervision for learning programming concepts.
Compared with augmentation-based baselines (SA-AFT, CTF-Code, and CodeFort), ProCURE maintains a clear advantage in CCScore while preserving competitive task accuracy. SA-AFT achieves the highest Pass@5 on Mistral-7B and StarCoder-7B, indicating that diversity-oriented code transformations mainly improve sampling-based success rates rather than concept-level generalization. Among the baselines, CodeFort yields the strongest CCScore improvements, reflecting the benefit of robustness-oriented adversarial perturbations that expose vulnerabilities related to insufficient concept understanding. In contrast, CTF-Code attains the lowest CCScore gains, as its instruction-level counterfactual augmentation provides limited supervision for learning fine-grained programming concepts.

\noindent \textbf{Ablation Analysis.}
Table~\ref{tab:ablation_avg} reports ablation results averaged over four models.
Removing both concept-sensitive loss terms, which reduces training to standard fine-tuning on the constructed dataset, leads to the largest performance degradation, with clear drops in pass@1 and pass@5 and a substantial decrease in CCScore.
This indicates that data augmentation alone is insufficient to capture programming concept variations.
When removing individual components, excluding the token-level loss $\mathcal{L}_{\text{token}}$ causes a more pronounced decline than excluding the attention-guided loss $\mathcal{L}_{\text{attn}}$, particularly in CCScore.
This suggests that explicit supervision on concept-altered tokens is the primary driver of concept understanding, while attention regularization provides complementary gains.
Overall, both components are necessary, and their combination yields the most consistent improvements.

\section{Related Work}

Perturbation-based code augmentation applies diverse code-specific transformations to existing programs to construct augmented training data for robust fine-tuning. Early studies indicate that neural code models are highly sensitive to minor perturbations, such as variable renaming or syntactic reordering, despite preserved program semantics~\cite{yefet2020adversarial,bielik2020adversarial}. These observations have motivated various augmentation approaches based on standard transformations or adversarial code samples.
A primary line of work focuses on rule-based or heuristic code transformations. ReCode~\cite{wang2023recode} systematically investigates over thirty semantic-preserving transformations, revealing substantial performance degradation in code generation models. More recently, CodeFort~\cite{zhang2024codefort} introduces a comprehensive robustness training framework combining diverse perturbations with consistency-based objectives. Another line of work explores adversarial perturbations. Prior studies design attacks, ranging from identifier substitution to structure-aware perturbations, to expose model vulnerabilities~\cite{zhou2022adversarial,zhang2020generating,tian2021generating}. CodeAttack~\cite{jha2023codeattack} further demonstrates the efficacy of adversarial fine-tuning as a defense mechanism. Subsequent research continues to refine these methods by leveraging syntactic or semantic structures to guide adversarial sample generation~\cite{chen2023evaluating}, with recent work extending this to adversarial perturbations specifically targeting code task descriptions~\cite{luo2025success}.

Crucially, Hooda et al.~\cite{hooda2024large} identify a ``programming concept understanding gap" in LLMs, highlighting the need for effective augmentation and fine-tuning strategies. However, existing methods are not tailored to programming concepts, limiting their ability to capture deep semantic logic. To address this, we propose a novel fine-tuning approach to enhance programming concept understanding.

\section{Conclusion}
We propose \textsc{ProCURE}, a concept-aware consistency learning framework that enhances LLMs’ understanding of programming concepts through concept-oriented code augmentation and concept-sensitive consistency learning. Extensive experiments on four open-source LLMs across three code generation benchmarks show that ProCURE improves CCScore by an average of 17.9 points, demonstrating its effectiveness in addressing the programming concept understanding gap.
\section*{Acknowledgments}
This work was supported by the 2025 Purple Mountain Talent Program for High-Level Innovative and Entrepreneurial Talents under the project “Theoretical Research on Trustworthy Evaluation of Intelligent Software” (Grant No. E6430219G8). This work was also supported by the National Natural Science Foundation of China (Grant Nos. U23B2041 and U24A6009), and the Joint Research Center for System Security, Tsinghua University (Institute for Network Sciences and Cyberspace)–Science City (Guangzhou) Digital Technology Group Co., Ltd.
\bibliographystyle{named}
\bibliography{ijcai26}

\end{document}